\documentclass[sn-mathphys,Numbered,iicol]{sn-jnl}% Math and Physical Sciences Reference Style
\usepackage{tikz}
\usepackage{tikz-3dplot}
\usepackage{pgfplots}
\usepackage{datatool}
\pgfplotsset{compat=1.15}
\usetikzlibrary{matrix}
\usepgfplotslibrary{units}
\usepgfplotslibrary{groupplots}
\usetikzlibrary{positioning,arrows.meta,calc}
\usetikzlibrary{angles, quotes, intersections, patterns}
\usetikzlibrary{decorations.pathmorphing}
\usepgfplotslibrary{colorbrewer} 
\usetikzlibrary{pgfplots.colorbrewer}
\usetikzlibrary{external}
\usepackage{bm} % amsfonts
\usepackage{siunitx}

\usepackage{graphicx}%
\usepackage{multirow}%
\usepackage{amsmath,amssymb,amsfonts}%
\usepackage{amsthm}%
\usepackage{mathrsfs}%
\usepackage[title]{appendix}%
\usepackage{xcolor}%
\usepackage{textcomp}%
\usepackage{manyfoot}%
\usepackage{booktabs}%
\usepackage{algorithm}%
\usepackage{algorithmicx}%
\usepackage{algpseudocode}%
\usepackage{listings}%
\newcommand{\norm}[1]{\left\lVert#1\right\rVert}
\newcommand{\nmse}{\operatorname{NMSE}}

\theoremstyle{thmstyleone}%
\theoremstyle{thmstyletwo}%

\theoremstyle{thmstylethree}%

\begin{document}

\title[Physics-Informed Neural Network for the Volumetric Sound field Reconstruction of Speech Signals]{Physics-Informed Neural Network for Volumetric Sound field Reconstruction of Speech Signals}


\author*[1]{\fnm{Marco} \sur{Olivieri}}\email{marco1.olivieri@polimi.it, xenoka@dtu.dk}
\author*[2]{\fnm{Xenofon} \sur{Karakonstantis}} %\email{xenoka@dtu.dk}
\author[1]{\fnm{Mirco} \sur{Pezzoli}} %\email{mirco.pezzoli@polimi.it}
\author[1]{\fnm{Fabio} \sur{Antonacci}} %\email{fabio.antonacci@polimi}
\author[1]{\fnm{Augusto} \sur{Sarti}} %\email{augusto.sarti@polimi.it}
\author[2]{\fnm{Efren} \sur{Fernandez-Grande}} %\email{efgr@dtu.dk}

\affil[1]{\orgdiv{Dipartimento di Elettronica, Informatica e Bioingegneria}, \orgname{Politecnico di Milano}} %, \orgaddress{\street{Via Ponzio 34/5}, \city{Milan}, \postcode{20133}, \country{Italy}}}

\affil[2]{\orgdiv{Department of Electrical \& Photonics Engineering}, \orgname{Technical University of Denmark}} %, \orgaddress{\street{Street}, \city{City}, \postcode{10587}, \state{State}, \country{Country}}}


\abstract{
Recent developments in acoustic signal processing have seen the integration of deep learning methodologies, alongside the continued prominence of classical wave expansion-based approaches, particularly in sound field reconstruction. Physics-Informed Neural Networks (PINNs) have emerged as a novel framework, bridging the gap between data-driven and model-based techniques for addressing physical phenomena governed by partial differential equations. This paper introduces a PINN-based approach for the recovery of arbitrary volumetric acoustic fields. The network incorporates the wave equation to impose a regularization on signal reconstruction in the time domain. This methodology enables the network to learn the underlying physics of sound propagation and allows for the complete characterization of the sound field based on a limited set of observations. The proposed method's efficacy is validated through experiments involving speech signals in a real-world environment, considering varying numbers of available measurements. Moreover, a comparative analysis is undertaken against state-of-the-art frequency-domain and time-domain reconstruction methods from existing literature, highlighting the increased accuracy across the various measurement configurations. 
%Notably, the proposed model demonstrates the ability to reconstruct the flow of acoustic energy as opposed to the baseline.
}

%%================================%%
%% Sample for structured abstract %%
%%================================%%

% \abstract{\textbf{Purpose:} The abstract serves both as a general introduction to the topic and as a brief, non-technical summary of the main results and their implications. The abstract must not include subheadings (unless expressly permitted in the journal's Instructions to Authors), equations or citations. As a guide the abstract should not exceed 200 words. Most journals do not set a hard limit however authors are advised to check the author instructions for the journal they are submitting to.
% 




\maketitle

%! TEX root=main
\section{Introduction}\label{sec:introduction}
% \begin{itemize}
    % \item Sound field reconstruction applications
    % \item Classes of sound field reconstruction methods
    % \item Brief description of a work for different classes
    % \item Parametric
    % \item Non-parametric/basis-function based
    % \item Deep leaning (generative models)
    % \item Physics-informed (Forum Acusticum papers)
    % \item Proposed method
    % \item Results
% \end{itemize}
Sound field reconstruction (SFR) is one of the main challenges in the literature of acoustic signal processing. 
It consists in the estimation of the acoustic %pressure 
field over an extended area using a limited set of multichannel acquisitions, typically obtained with distributed microphones or arrays.
SFR is of paramount importance in augmented and virtual reality applications, where the user experience hinges on immersive audio environments \cite{vorlander2015virtual}.
A complete acquisition of an acoustic environment i.e., satisfying the Nyquist–Shannon sampling theorem in the audible range, imposes an extremely dense sampling of the space (lower than $\SI{1}{\centi\meter}$). 
This condition limits the practical implementation of such applications raising the interest in SFR for reducing the number of required measurements. 
Beyond its critical role in augmented and virtual reality applications, SFR finds application across diverse problems including sound field control \cite{tohyama2000fundamentals}, separation \cite{pezzoli2021ray} and localization \cite{cobos2017survey}.

Several solutions to SFR have been introduced in the literature, predominantly tackling the reconstruction in terms of an inverse problem. 
The acquisition of the sound field relies on a few selected measurement points, from which the field at other locations can be estimated, via interpolation or extrapolation.
%on few selected measurement points, while the estimation in ``unknown'' arbitrary locations has the goal of either interpolate or extrapolate the information.
In the literature, we can identify three main classes of SFR methodologies: non-parametric or expansion-based \cite{koyama2019sparse, ueno2018kernel, ueno2017sound, pezzoli2022sparsity}, parametric \cite{pezzoli2018reconstruction, pulkki2018parametric, del2011generating}, and Deep Learning (DL) \cite{lluis2020sound, kristoffersen2021deep, pezzoli2022deep}.

%Most SFR models adopt a representation of the sound field using expansions of basis functions. 
%In general, such non-parametric methods exploit the well-known solutions of the wave equations i.e., plane wave or spherical waves, in order to model the acoustic field. 
%The measured data is projected onto a linear combination of basis functions aiming at numerically estimate the sound field.
%Other solutions employ the modal representation \cite{das2021room, haneda1999common} or the equivalent source method \cite{lee2017use} (ESM) to describe the sound field. 

In order to estimate the coefficients of the expansion-based representation, most techniques rely on compressed sensing principles \cite{donoho2006compressed} exploiting a priori assumptions on the sparsity of the data either in time, space or frequency. 
In \cite{koyama2019sparse}, the equivalent source method \cite{lee2017use} (ESM) is adopted in order to model the direct sound of sound sources through a sparse set of Green's function, while reverberation is described by means of a dictionary of plane waves and a low-rank residual. 
The solution in \cite{koyama2019sparse} is found using sparsity-based optimization. 
A similar strategy, defined in the spherical harmonics domain, is proposed in \cite{pezzoli2022sparsity}.
The authors of \cite{antonello2017room}, instead, exploit the ESM and the sparsity of room impulse response (RIR) signals in the time domain (TESM) to represent the far-field sound field in a source-free volume.
In contrast, \cite{caviedes2023spatio} further extends this approach to dynamically regularize the solution based on the reverberation characteristics present in a room.
More recently, the utilization of dictionary learning featuring more spatially constrained representations of the acoustic environment has demonstrated efficacy in achieving generalization across diverse sound fields. \cite{hahmann2021spatial}

Another popular strategy relies on the so-called kernel ridge regression (KRR) \cite{vovk2013kernel} which provides a least-square-based solution to the reconstruction.
One advantage of KRR lies in the lightweight implementation provided by linear filtering.
In \cite{ueno2018kernel} a sound field interpolation strategy based on KRR has been introduced and has since been extended to multiple arrays also \cite{duran2023room}. 
This method exploits the expansion into infinite sum plane wave functions in order to force the interpolation to satisfy the Helmholtz equation, inherently imposing a physical prior to the solution. 
Variations to the kernels function has been proposed, including prior information on the source direction \cite{ribeiro2022region}, constraints on the reciprocity principle \cite{ribeiro2020kernel} or mixed models for the reverberation \cite{ribeiro2023kernel}.
In \cite{caviedes2021gaussian} the authors evaluate different kernel definitions for KRR-based sound field reconstruction analyzing the methodology in the framework of Gaussian process regression. 

Differently from previously described expansion-based solutions, parametric approaches \cite{9795678,pezzoli2020parametric, pezzoli2018reconstruction, pulkki2018parametric, thiergart2013geometry} are not targeted to numerically reconstruct the sound field. 
As a matter of fact, parametric techniques rely on simplified signal models that describe the sound field by means of a few parameters e.g., the source location, its direct signal and the diffuse field to convey a perceptually convincing reconstruction.
Typically, the parameters are estimated from the microphone signals using beamforming \cite{gannot2017consolidated} or other linear filtering \cite{mignot2013low, jin2015theory} and the sound field reconstruction is provided at the user through the adopted signal model.

% Recently, data-driven approaches \cite{} emerged as an effective alternative for SFR, in particular deep neural network (DNN) \cite{}. 
Following the success of deep neural network (DNN) in several problems of acoustics \cite{olivieri2021physics, olivieri2021audio, olivieri2023real, karakonstantis2021sound}, deep learning solutions became a popular approach for SFR. 
Seminal work in \cite{lluis2020sound} adopted a convolutional neural network (CNN) for the reconstruction of room transfer functions. 
The main limitation in applying classical DNN in SFR is represented by the large amount of data required for the training of the model. 
This typically translates into a reconstruction limited to low frequencies or lacking of generalization to different data \cite{lluis2020sound}. 
In order to overcome such limitations two main strategies can be identified in the literature. 
On the one side, generative models \cite{fernandez2023generative, karakonstantis2023generative, miotello2023reconstruction} have been introduced to improve the ability of the network in reconstructing meaningful sound fields. 
On the other side, domain knowledge has been employed to help the neural network to follow the physics of the acoustic phenomena.

In \cite{fernandez2023generative} a generative adversarial network (GAN) has been trained to solve the SFR problem. 
Different GAN designs have been tested in \cite{fernandez2023generative} and the results revealed improved reconstruction and bandwidth extension with respect to the customary plane wave decomposition both on simulated and real data sets. 
More recently, GANs have been combined with the physical prior given by plane wave expansion in order to exploit acoustic models in the generation. 
In particular, in \cite{karakonstantis2023generative} the authors use a generator network to compute the coefficients of the plane wave expansion whose reconstruction is evaluated by the discriminator network for the adversarial training. 
%Similar approach has been employed in \cite{}, where the normalizing flow paradigm is adopted. 
Differently from the aforementioned techniques that work in the frequency domain, a time domain model has been introduced in \cite{pezzoli2022deep}, where a deep-prior paradigm \cite{ulyanov2018deep} for the generation of RIR is proposed. 
The deep prior technique employs the structure of CNN as a regularizor of the SFR solution, hence does not require a training data set, but the network is rather trained using pre-element training.
Although very accurate also when a small set of pressure data is available, the applicability of such techniques strongly depends on the training set with limitations in the generalization to different rooms, e.g., non-rectangular rooms with complex boundary conditions~\cite{lluis2020sound, kristoffersen2021deep}.

More recently, the use of Physics-Informed Neural Networks (PINNs) has been investigated to bridge the gap between model-based solutions, which are constrained by the underlying modeling assumptions, and data-driven methods, whose solutions largely depend on training data.
PINNs aim to regularize the estimates of a neural network to follow a given partial differential equation (PDE) governing the system under analysis, thus providing physics-informed solutions.
In the context of SFR, authors in~\cite{shigemi2022physics} proposed a CNN whose loss function penalizes deviation from the Helmholtz equation~\cite{williams1999fourier}.
The computation of the desired PDE is computed through numerical methods to avoid the spatial discretization of the pressure distribution.
More computationally efficient approaches take advantage of the automatic differentiation framework~\cite{baydin2018automatic} underlying the training procedure of neural networks~\cite{raissi2019physics}.
%As a matter of fact, the literature of PINN-based method~\cite{raissi2019physics, cuomo2022scientific, karniadakis2021physics} proved how the MLP architecture (Section~\ref{sec:ANN}) with hyperbolic tangent~\eqref{eq:hyperbolicFunction} activation functions is a suitable option for inferring solutions to PDEs, thus obtaining physics-informed surrogate models that are fully differentiable with respect to all input coordinates.

Different PINN approaches for the reconstruction of sound fields have been presented in~\cite{pezzoli2023implicit, karakonstantis2023room, karakonstantis2024room}.
Inspired by the SIREN network~\cite{sitzmann2020implicit}, %the authors
different studies~\cite{pezzoli2023implicit, karakonstantis2023room, karakonstantis2024room} proposed a PINN-based approach for the reconstruction of sound fields to efficiently recover the RIRs in time domain starting from a small number of available observations. Moreover, in \cite{karakonstantis2023room, karakonstantis2024room}, PINNs are employed to comprehensively characterize all acoustic quantities in the sound field. This includes the inference of the pressure field, particle velocity field, and sound intensity flows, by exploiting the automatic differentiation principle of neural network.
% Although these methods showed accurate solutions both with simulated and real data, their applicability is limited to the reconstruction of the early part of RIRs~\cite{pezzoli2023implicit, karakonstantis2023room} discarding the reverberant contribution of the room.
As do the majority of DL-based methods, they focus on the reconstruction of height-invariant scenarios, typically by limiting to the case with sources and the emitters placed in the same plane.
% 2D scenarios, thus requiring the sources and the emitters placed in the same plane. 

In this manuscript, we aim at extending the methodology introduced in~\cite{pezzoli2023implicit, karakonstantis2023room} to reconstruct arbitrarily sound fields in 3D target regions starting from different sets of available recordings.
We exploit the potential of DL estimates with the regularization provided by the physical knowledge of acoustic wave propagation.
%In particular, we propose a MLP architecture that from the desired space and time domain in input is able to recover the time-domain pressure signals of the acoustic field.
%Hence, estimating all the acoustic pressure contributions of the space.
Different from considering only the RIRs, we consider speech signals in real environments, with a relevant step towards practical application scenarios.
Specifically, the devised DNN is driven by the available measurements and constrained to satisfy the physical PDE of the wave equation.
Moreover, we directly reconstruct the signals in time-domain, thus fully characterizing the sound pressure at the desired locations.
The reconstruction of the proposed method has been evaluated using real data, i.e., MeshRIR dataset~\cite{koyama2021meshrir}, coming from acoustically treated rooms.
Starting from different random subsets of available observations, results show improved reconstruction performances with respect to frequency-domain~\cite{ueno2018kernel} and time-domain~\cite{antonello2017room} state-of-the-art approaches.
The promising results of the devised method prove how PINN represents an appealing solution to generalize the reconstruction of the sound field towards practical scenarios, thanks to the advantages of the acoustic physical priors and the potential of DL strategies to infer representations from real data.
Furthermore, %exploiting the automatic differentiation principle of neural network, 
we show the computation of the reconstructed intensity field in the desired volumetric region.

The rest of the paper is organized as follows.
In Sec.~\ref{sec:data_model_problem_formulation}, we define the data model~\ref{subsec:data_model} and the problem formulation~\ref{subsec:problem_formulation}.
Sec.~\ref{sec:proposed_method} presents the proposed method based on PINN and details information about the network architecture and training procedure are described in Sec.~\ref{subsec6:pinn_architecture}.
Results are reported in Sec.~\ref{sec:results} along with the comparisons with respect to state-of-the-art SFR methods.
Finally, Sec.~\ref{sec:conclusions} draw some conclusions and future developments.
%! TEX root=main
\section{Data model and problem formulation}\label{sec:data_model_problem_formulation}

% \begin{figure}[t]%
%     \centering
%     \input{tikz/cartesian_coordinate.tikz}
%     %\includegraphics[width=0.5\linewidth]{tikzfigures/cartesian_coordinate.pdf}
%     \caption{Figure example tikz.}
%     \label{fig:example}
% \end{figure}

% ----------------------------------------------
\subsection{Data model}\label{subsec:data_model}
Let us consider an acoustic source located at an arbitrary position $\bm{r}_\mathrm{s}=[x_\mathrm{s}, y_\mathrm{s}, z_\mathrm{s}]^T$ and a set $\mathcal{M}$ of $M$ measurements acquiring the generated sound field at positions $\bm{r}_\mathrm{m}=[x_\mathrm{m}, y_\mathrm{m}, z_\mathrm{m}]^T$ with $\mathrm{m}=1,\dots, M$.
Under the assumption of a Linear Time-Invariant (LTI) acoustic system and in the absence of noise, the acoustic pressure acquired by the $\mathrm{m}^\mathrm{th}$ sensor can be expressed as
\begin{equation}\label{eq6:mic_from_rir}
    p(\bm{r}_\mathrm{m}, t) = h_{\mathrm{m}, \mathrm{s}}(t) * s(t),
\end{equation}
where symbol $*$ denotes the linear convolution operator, $p(\bm{r}_\mathrm{m}, t)$ is the time-domain sound pressure measured at location $\bm{r}_\mathrm{m}$, $s(t)$ is the signal emitted by the source, and $h_{\mathrm{m}, \mathrm{s}}(t)$ is the Room Impulse Response (RIR) function that describes the %spatio-temporal propagation of the sound 
transfer path of sound from the source in $\bm{r}_\mathrm{s}$ to the receiver at $\bm{r}_\mathrm{m}$.
Notice that, due to the LTI assumption and in ideal conditions with unbounded domain, the RIR in~\eqref{eq6:mic_from_rir} is the solution of the inhomogeneous wave equation% ~\eqref{eq:inhomWaveEquation}  and it can be expressed in frequency-domain by the well-known Green's function $G(\bm{r}_\mathrm{m}|\bm{r}_\mathrm{s}, \omega)$~\eqref{eq:greenFunction}.
\begin{equation}
    \left( \nabla^2 - \frac{1}{c^2}\frac{\partial^2}{\partial t^2} \right) p(\bm{r}_\mathrm{m}, t) = s(\bm{r}_\mathrm{s},t),
\end{equation}
where $c$ is the speed of sound, $\nabla^2$ is the Laplacian operator, and the source term $s$ is considered to be dependent on position $\bm{r}_\mathrm{s}$ in this instance.

Obtaining a RIR experimentally typically entails the use of a receiver-emitter combination, the first of which records the variation in sound pressure over time produced by the latter through the principle of transduction (i.e. microphone and loudspeaker), as well as a post-processing stage, which requires the deconvolution of the acquired signal with respect to the emitted source signal~\cite{stan2002comparison, farina2007advancements}.
%a specific technique available in the literature based on the signal adopted for the measurement~\cite{stan2002comparison, farina2007advancements}.
In practical applications, the sound pressure is acquired at $M$ discrete sensor positions and is commonly organized in a $N\times M$ matrix defined as
\begin{equation}\label{eq6:pressure_matrix}
    \mathbf{P} = \tilde{p}(\bm{r}_\mathrm{m}, t_\mathrm{n}) = [\mathbf{p}_\mathrm{1}, \dots, \mathbf{p}_\mathrm{m}, \dots, \mathbf{p}_\mathrm{M}],
\end{equation}
where $\mathbf{p}_\mathrm{m} \in \mathbb{R}^{N\times 1}$ is the vector containing the $N$-length sampled pressure~\eqref{eq6:mic_from_rir} at position $\bm{r}_\mathrm{m}$ and time $t_n \subset t$.

\subsection{Problem formulation}\label{subsec:problem_formulation}

This section addresses the challenge of determining a function that accurately represents the intrinsic pressure field $p(\bm{r}, t)$ based on a restricted set of observations $\tilde{p}(\bm{r}_\mathrm{m}, t_\mathrm{n})$. With $\mathcal{M}$ denoting the maximum number of sensors or measurements, the objective is delineated by
\begin{equation}\label{eq6:pressure_minimisation_physical}
    \begin{split}
        \hat{\bm{\beta}}_{\mathrm{opt}} &= \underset{\hat{ \bm{\beta}}}{\arg\min} \left( |p( \bm{\beta}, \bm{r}_\mathrm{m}, t_\mathrm{n}) - \tilde{p}(\bm{r}_\mathrm{m}, t_\mathrm{n})|^2 \right) \\
        & \text{s.t.} \quad \forall \mathrm{m} \in \mathcal{M}\\
        & \left( \nabla^2 - \frac{1}{c^2}\frac{\partial^2}{\partial t^2} \right) p( \bm{\beta}, \bm{r}, t) = 0,
    \end{split}
\end{equation}
where $ \bm{\beta}$ are the model parameters. Thus, the aim is to minimize the discrepancy between the estimated pressure function and the observed data while adhering to the wave equation constraint either implicitly (analytic basis function expansions) or explicitly. By limiting the set of measurements to a subset of the original set, denoted as $\tilde{\mathcal{M}}$ (where $\tilde{\mathcal{M}} \subseteq \mathcal{M}$ and $|\tilde{\mathcal{M}}| = \tilde{M} < M$), we explore the sensitivity of modeling this function using various array configurations. Numerous strategies for computing the unknown pressure function exist, ranging from implicit solutions to the homogeneous wave equation~\cite{antonello2017room, damiano2021soundfield, zea2019compressed, ueno2018kernel, fernandez2021reconstruction} to those focusing on data fidelity~\cite{pezzoli2018reconstruction, pezzoli2022deep, lluis2020sound}, and even combinations of both~\cite{karakonstantis2023generative}. However, explicitly incorporating the wave equation as a constraint follows recent developments in the literature.

%well-known Green's function $G(\bm{r}_\mathrm{m}|\bm{r}_\mathrm{s}, \omega)$, namely
% \begin{equation}\label{eq:inhomHelmholtzEquationSolution}
%     \nabla^2 G(\bm{r}_\mathrm{m}|\mathrm{s}, \omega) + k^2 G(\bm{r}|\bm{r'}, \omega) = -\delta(\bm{r}-\bm{r'}),
% \end{equation}
% where $\nabla^2 = \nabla \cdot (\nabla)$ is the Laplace operator~\cite{morse1954methods} and $\omega = 2\pi f$ is the natural frequency for the frequency $f$, $k$ represents the wavenumber and it is defined as $k = \frac{\omega}{c}$, with $c=\SI{343}{\meter/\second}$ the speed of sound in air, and $j$ the imaginary unit, thus $\sqrt{-1}=j$.
%\input{chapters/related_work}
% ----------------------------------------------
\section{Proposed method}\label{sec:proposed_method}

This study formulates the constrained optimization problem in~\eqref{eq6:pressure_minimisation_physical} using a PINN, thus adopting a neural network $\mathcal{N}(\cdot)$ that takes as input the signal domain, i.e., the scalar time and position values where the pressure is to be evaluated, and provides as output an estimate of the pressure field. This implies that the function of pressure is given by 
% constrained by the physical solution of the wave equation as in~\eqref{eq6:pressure_minimisation_physical}, namely
\begin{equation}\label{eq6:network_generator}
    p(\bm{r}, t) = \mathcal{N}(\bm{\Theta},\bm{r}, t),
\end{equation}
where $\bm{\Theta}$ are the neural network parameters.
Therefore, we can rewrite the optimization in~\eqref{eq6:pressure_minimisation_physical} in order to find the optimal weights $\bm{\Theta}_{\mathrm{opt}}$ that parameterize the network as
\begin{equation}\label{eq6:theta_pressure_minimisation_physical}
    \begin{split}
        \bm{\Theta}_{\mathrm{opt}} &= \underset{\bm{\Theta}}{\arg\min}\ |\mathcal{N}(\bm{\Theta},\bm{r}_{\mathrm{m}},  t) - \tilde{p}(\bm{r}_\mathrm{m}, t)|^2 \\
        & \text{s.t.} \quad \forall \ \mathrm{m} \in \mathcal{M}\\
        & \left( \nabla^2 - \frac{1}{c^2}\frac{\partial^2}{\partial t^2} \right) \mathcal{N}(\bm{\Theta}, \bm{r}, t) = 0.
    \end{split}
\end{equation}
Notice that, although the data-fidelity term in \eqref{eq6:theta_pressure_minimisation_physical} is computed only with the available observations in $\mathcal{M}$, the physical regularization given by the wave equation is applied for all positions $\bm{r}$ in the domain.

Given that the optimal weights $\bm{\Theta}_{\mathrm{opt}}$ have been obtained, one can obtain the particle velocity via Euler's equation of motion (a result of the conservation of momentum) \cite{karakonstantis2024room}
\begin{equation}\label{eq:p_velocity}
    \begin{split}
        \bm{u}(r,t) &= -\frac{1}{\rho}\int_{t_0}^{t} \nabla p(\bm{r}, t) \partial t \\ &= -\frac{1}{\rho}\int_{t_0}^{t} \nabla \mathcal{N}(\bm{\Theta}_{\mathrm{opt}}, \bm{r}, t) \partial t,
    \end{split}
\end{equation}
between time $t_0$ and $t$, where the gradient $\nabla \mathcal{N}(\bm{\Theta}_{\mathrm{opt}}, \bm{r}, t)$ is obtained via automatic differentiation, while $\rho$ represents the density of the fluid medium. 

Together with the pressure and the particle velocity, the instantaneous intensity field is obtained from the product 
\begin{equation}\label{eq:insta_intensity}
    \bm{i}(\bm{r}, t) =  \bm{u}(\bm{r}, t) \cdot p(\bm{r}, t),
\end{equation}
which allows for a complete characterization of the reconstructed sound field anywhere in the domain of interest.

%---------------------------------------------------
\subsection{Neural Network architecture description}\label{subsec6:pinn_architecture}
In the following, we describe in detail the proposed model $\mathcal{N}(\cdot)$, along with the definition of the adopted loss function and training procedure.

\vspace*{5mm}
\noindent \textbf{SIREN-inspired neural network}\\
Inspired by recent works presented in~\cite{pezzoli2023implicit, karakonstantis2023room}, which proved the ability of SIREN-like architectures~\cite{sitzmann2020implicit} to learn an implicit representation of time-domain signals, thanks to the adoption of sinusoidal nonlinearities, we propose a similar PINN-SIREN architecture to recover the audio signals in a 3D volume.

The devised architecture parameterized by the learnable weights $\bm{\Theta}$ and the input $\bm{\mathsf{x}}$ has the structure of a multilayer perceptron (MLP)~\cite{hornik1989multilayer} with $L$ layers, whose structure can be expressed as
\begin{equation}\label{eq6:mlp_structure}
    \mathcal{N}(\bm{\Theta}, \bm{\mathsf{x}}) = (\Phi_L \circ \Phi_{L-1} \circ \dots \circ \Phi_{1})(\bm{\mathsf{x}}).
\end{equation}
As for the SIREN network~\cite{sitzmann2020implicit}, the $\mathrm{i}^{\mathrm{th}}$ layer is characterized by a sinusoidal activation function, namely
\begin{equation}\label{eq6:siren_layer}
    \Phi_\mathrm{i}(\bm{\mathsf{x}}_\mathrm{i}) = \sin \left( \omega_0 \bm{\mathsf{x}}_\mathrm{i}^T \bm{\Theta}_\mathrm{i} + \bm{\mathsf{b}}_\mathrm{i} \right),
\end{equation}
where $\bm{\mathsf{x}}_\mathrm{i}$, $\bm{\Theta}_\mathrm{i}$, and $\bm{\mathsf{b}}_\mathrm{i}$ are the input vector, the weights and the biases of the $\mathrm{i}^{\mathrm{th}}$ layer, respectively, and $\omega_0$ represents an initialization parameter~\cite{sitzmann2020implicit}.

%Authors in~\cite{sitzmann2020implicit} proved how the sinusoidal nonlinearity provides an effective implicit representation of the time-domain signals.
% showing , the authors show how a two-layers SIREN can be related to a discrete cosine transform (DCT) of the signal.
    
\vspace*{5mm}
\noindent \textbf{Input/Output Data}\\
The network input is the $N \times M\times 4$ tensor representing the combination of all the possible points of the domain, i.e., time samples and 3D spatial coordinates, of the desired microphones, namely
\begin{equation}\label{eq6:network_input}
    [t_{\mathrm{n}}, x_{\mathrm{m}}, y_{\mathrm{m}}, z_{\mathrm{m}}] \quad \mathrm{n} \in N, \mathrm{m} \in M.
\end{equation}
On the other hand, the output of the network is the scalar pressure values of $p(t_{\mathrm{n}}, x_{\mathrm{m}}, y_{\mathrm{m}}, z_{\mathrm{m}})$ which can then be reorganized into the $N$-length time-domain pressure signals as given by Eq. \eqref{eq6:pressure_matrix}.

It is worth noticing that authors in~\cite{sitzmann2020implicit} normalized the input of SIREN architecture, both time and space coordinates, in the range $[-1, 1]$ to converge towards a correct numerical solution.
Moreover, they showed how the initialization of the parameter $\omega_0$ in~\eqref{eq6:siren_layer} for different values spans multiple periods of the sinusoidal activations over $[-1, 1]$ and how it affects the frequency content of the reconstructed signals.

In this work, differently from the experiments presented in~\cite{sitzmann2020implicit} that consider a single time-domain signal, the devised network structure is modified in order to deal with 4D input domain~\eqref{eq6:network_input}.
Moreover, although we consider the space components, i.e., $\mathrm{x}$, $\mathrm{y}$, and $\mathrm{z}$, of the input in the range $[-1, 1]$ as in~\cite{sitzmann2020implicit}, we increase the range of the time component in $[-100, 100]$.
Imposing such a different time range with respect to the space coordinates is equivalent to adding a constant multiplication term to the weights of the input layer of the network, as demonstrated in~\cite{sitzmann2020implicit}.
This different normalization enables the network to specialize the frequency contents in time and space, separately, thus accounting for the high temporal frequency contents of the speech signals and the low spatial frequency positions of the microphones.

\begin{figure}[!tb]
    \centering
    \includegraphics[width=\columnwidth]{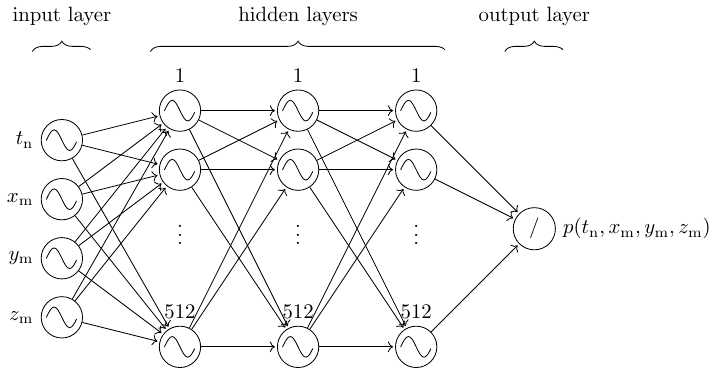}
    \caption{Schematic of the proposed MLP architecture. The time and spatial domain points in input, while the estimate of the acoustic pressure in output. The $3$ hidden layers with $512$ neurons are characterized by sinusoidal activation functions, while the last layer is linear.}
    \label{fig6:mlp_architecture}
\end{figure}

\vspace*{5mm}
\noindent \textbf{MLP structure}\\
The proposed architecture~\eqref{eq6:mlp_structure} is composed of $L = 5$ layers of $512$ neurons, as depicted in Figure~\ref{fig6:mlp_architecture}.
The first four layers are characterized by sinusoidal activation functions \eqref{eq6:siren_layer}, while the last layer is linear, thus achieving a total number of learnable parameters around $\SI{790000}{}$.
Moreover, the initialization frequency $\omega_0$ in~\eqref{eq6:siren_layer} is experimentally set to $0.5$ for the first layer, while $\omega_0 = 30$ for the hidden layers. Finally, due to the scaling of the input collocation points, we also scaled the speed of sound to reflect the scaled travel time during training.

\vspace*{5mm}
\noindent \textbf{Training procedure}\\
To imbue the neural network with the physical constraints outlined in the wave equation~\eqref{eq6:theta_pressure_minimisation_physical}, we employ the following loss function
\begin{equation}\label{eq6:pinn_loss_reconstruction}
    \begin{split}
        \mathcal{L} &= \frac{1}{\mathcal{M}} \sum_{\mathrm{m}\in \mathcal{M}} \norm{\hat{p}(\bm{r}_\mathrm{m}, t_\mathrm{n}) - \tilde{p}(\bm{r}_\mathrm{m}, t_\mathrm{n})}^2 \\
        & +
        \lambda \frac{1}{Q} \sum_{\mathrm{q} = 1}^{Q} \norm{\nabla^2 \Hat{p}(\bm{r}_\mathrm{q}, t) - \frac{1}{c^2} \frac{\partial^2}{\partial t^2} \Hat{p}(\bm{r}_\mathrm{q}, t) }^2.
     \end{split}
\end{equation}	
Here, $\Hat{p}$ and $\tilde{p}$ denote the network estimate and the measured sound pressure, respectively. The summation over $Q$ positions is performed to ensure the fulfilment of the wave equation in a batch-like setting. Essentially, Eq. \eqref{eq6:pinn_loss_reconstruction} represents the Lagrangian form of the constrained objective detailed in Eqs. \eqref{eq6:pressure_minimisation_physical} and \eqref{eq6:theta_pressure_minimisation_physical}. The parameter $\lambda$ in~\eqref{eq6:pinn_loss_reconstruction} serves to balance the contributions of the two terms in the loss function and has been maintained at a constant value of $1 \cdot 10^{-5}$ throughout this study.

We train the devised model for $\SI{3000}{}$ iterations on a single NVIDIA Titan RTX GPU with $\SI{24}{\giga\byte}$ of memory, and we adopt Adam optimizer~\cite{kingma2015adam} with learning rate equal to $5\cdot10^{-5}$ to compute the gradient descent algorithm.
%! TEX root=main
\section{Results}\label{sec:results}
In the following, we present the results of the %devised 
sound field reconstruction method considering different subsets of available observations.
Firstly, we provide the setup adopted and examples of acoustic field reconstruction in the desired 3D region.
Then, we compare the performance with respect to the Kernel interpolation method~\cite{ueno2018kernel} and the TESM~\cite{antonello2017room} approaches.

\begin{figure*}[!tb]
    \centering
    \includegraphics[width=0.8\textwidth]{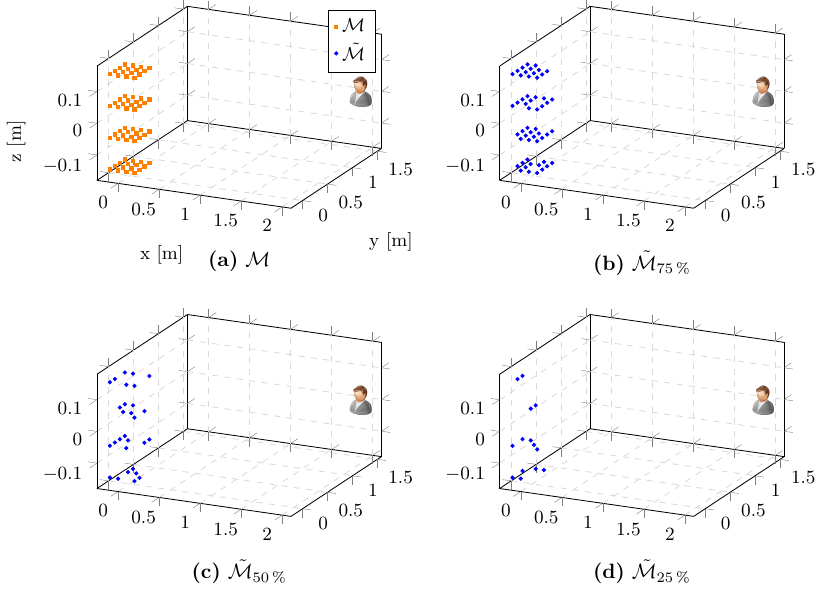}
    \caption{Setups adopted for the devised 3D sound field reconstruction method. The square-shaped orange marks are microphones in $\mathcal{M}$, while the circle-shaped blue marks represent the available observations $\tilde{\mathcal{M}}$. In (a) the $64$ positions of the acoustic field, while (b), (c), and (d) show the different setups with $\SI{75}{\percent}$, $\SI{50}{\percent}$, and $\SI{25}{\percent}$ of available microphones, respectively. The head icon represents the source position.}
    \label{fig6:mic_setup}
\end{figure*}

\vspace*{5mm}
\noindent \textbf{Setup}\\
We consider the MeshRIR dataset~\cite{koyama2021meshrir} to evaluate the devised method with real data.
This dataset collects RIRs inside a 3D region measured from a single source position placed outside the cuboidal space, as depicted in Figure~\ref{fig6:mic_setup}a.

The RIRs at each position have been measured with an omnidirectional microphone (Primo, EM272J) by recording the signal emitted with a loudspeaker (DIATONE, DS-7), as described in~\cite{koyama2021meshrir}.
The measurements have been conducted inside a room with dimensions $7 \times 6.4 \times 2.7~\si{\meter}$, reverberation time $\mathrm{T_{60}} = \SI{0.38}{\second}$, and temperature around $\SI{26.3}{\degreeCelsius}$, thus the estimated sound speed is $c = \SI{346.8}{\meter/\second}$.

We consider different subsets of the MeshRIR data to retrieve the acoustic field in $M=64$ positions arranged in an equally spaced grid of $4 \times 4 \times 4$ points, with inter-sensor distance $d=\SI{0.1}{\meter}$ in the $\mathrm{x}$, $\mathrm{y}$, and $\mathrm{z}$ dimension.
The desired speech signals are obtained according to \eqref{eq6:mic_from_rir} with the convolution operator between the measured RIRs and a clean speech recording, thus obtaining the time-domain signals sampled at $\SI{16000}{\hertz}$.
Moreover, due to the capacity limitations of the GPU memory, we consider signals with $N=800$ samples, and we recover the acoustic fields for different randomly selected windows of the speech signals.
Therefore, we collect the pressures in the matrix $\mathbf{P}\in\mathbb{R}^{800 \times 64}$~\eqref{eq6:pressure_matrix}, which represents also the ground truth of the reconstruction method.

%In the following, we aim at estimating $\Hat{\mathbf{P}}$ starting from a limited set of observations $\tilde{\mathcal{M}}$.
With reference to Figure~\ref{fig6:mic_setup}, we consider three different scenarios in which $3/4$, $1/2$, and $1/4$ of the total sensors $\mathcal{M}$ are available.
In the following, we will denote with $\tilde{\mathcal{M}}_{\SI{75}{\percent}}$, $\tilde{\mathcal{M}}_{\SI{50}{\percent}}$, and $\tilde{\mathcal{M}}_{\SI{25}{\percent}}$ the set of available observations corresponding to $48$, $32$, and $16$ randomly selected microphones, respectively.

Figure~\ref{fig6:mic_setup}a shows the set $\mathcal{M}$ of microphones corresponding to the desired $|\mathcal{M}| = 64$ sensors.
The different setups with $\tilde{\mathcal{M}}_{\SI{75}{\percent}}$, $\tilde{\mathcal{M}}_{\SI{50}{\percent}}$, and $\tilde{\mathcal{M}}_{\SI{25}{\percent}}$ of the available observations are depicted in Figure~\ref{fig6:mic_setup}b, Figure~\ref{fig6:mic_setup}c, and Figure~\ref{fig6:mic_setup}d, respectively.
Aiming to recover the time-domain pressure signals in the $64$ positions of the 3D region from different available observations, we denote the pressure estimations with different subscripts to identify the dimension of observation subset in input, namely $\Hat{\mathbf{P}}_{\SI{75}{\percent}}$, $\Hat{\mathbf{P}}_{\SI{50}{\percent}}$, and $\Hat{\mathbf{P}}_{\SI{25}{\percent}}$.

\begin{figure*}[!tb]
    \centering
    \includegraphics[width=0.9\textwidth]{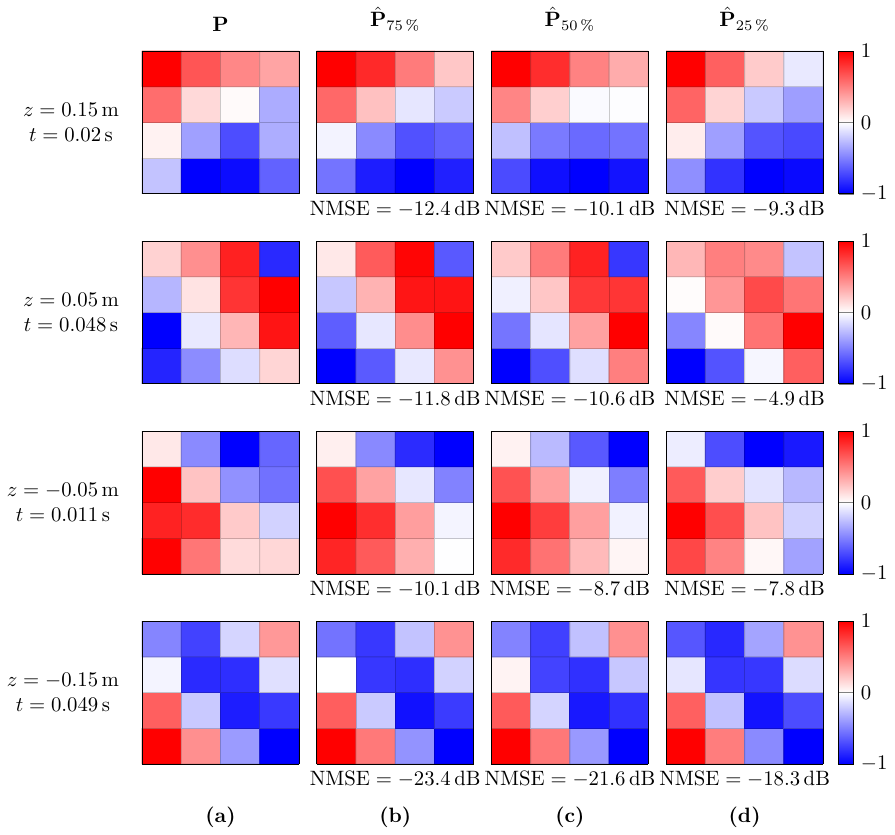}
    \caption{Reconstruction examples at different time snapshots computed from different subsets of observations. The acoustic field is depicted in the $\mathrm{xy}$-plane at the different elevations of the desired grid points. Column (a) show the ground truth of the sound field. In columns (b), (c), and (d) the reconstructions computed from $\SI{75}{\percent}$, $\SI{50}{\percent}$, and $\SI{25}{\percent}$ of microphones are depicted along with the $\nmse$ value with respect to the recovered acoustic field and the ground truth in the plane. The sound field in each row is normalized in the range $[-1,1]$.}
    \label{fig6:pressure_reconstruction_examples}
\end{figure*}

\vspace*{5mm}
\noindent \textbf{Evaluation metrics}\\
The performance of the reconstructions is assessed in terms of the Normalized Mean Square Error ($\mathrm{NMSE}$), expressed in decibels, between the estimated data and the reference acoustic field as
\begin{equation}\label{eq6:nmse_tot}
    \nmse(\Hat{\mathbf{P}}, \mathbf{P}) = 10\log_{10} \frac{1}{M} \sum_{\mathrm{m}=1}^M \frac{\norm{\Hat{\mathbf{p}}_\mathrm{m} - \mathbf{p}_\mathrm{m}}^2}{\norm{\mathbf{p}_\mathrm{m}}^2}.
\end{equation}
Moreover, to evaluate the reconstruction accuracy in the positions that do not belong to the available observations, i.e., $\mathrm{m}\notin\tilde{\mathcal{M}}$, we define with
\begin{equation}\label{eq6:nmse_val}
    \nmse_\mathrm{VAL} = 10\log_{10} \frac{1}{M-\tilde{M}} \sum_{\mathrm{m}\notin\tilde{\mathcal{M}}} \frac{\norm{\Hat{\mathbf{p}}_\mathrm{m} - \mathbf{p}_\mathrm{m}}^2}{\norm{\mathbf{p}_\mathrm{m}}^2}
\end{equation}
the $\nmse$ of the missing channels, % in $\tilde{{\mathbf{P}}}$~\eqref{eq6:pressure_incomplete}, 
and with
\begin{equation}\label{eq6:nmse_sig}
    \nmse_\mathrm{SIG} = 10\log_{10} \frac{1}{\tilde{M}} \sum_{\tilde{\mathrm{m}}\in\tilde{\mathcal{M}}} \frac{\norm{\Hat{\mathbf{p}}_{\tilde{{\mathrm{m}}}} - \mathbf{p}_{\tilde{{\mathrm{m}}}}}^2}{\norm{\mathbf{p}_{\tilde{{\mathrm{m}}}}}^2}
\end{equation}
we consider the error between the available observation and the fitted data.

Notice that, we are interested in maximizing the performance for $\nmse_\mathrm{VAL}$ to obtain the optimal reconstruction of the acoustic field in the desired 3D region.

\vspace*{5mm}
\noindent \textbf{Validation}\\
In order to assess the effectiveness of the proposed physics-informed methodology, we evaluate the reconstruction performance in ten different randomly selected time windows of the speech signals.
Therefore, we evaluate $\Hat{\mathbf{P}}$ multiple times considering all the time windows and the three different scenarios of microphone setups.

In Figure~\ref{fig6:pressure_reconstruction_examples}, we show different examples of acoustic field reconstruction for different time snapshots.
The first column depicts the ground truth $\mathbf{P}$ of the sound field, while the second, third, and last columns show the reconstructions $\Hat{\mathbf{P}}_{\SI{75}{\percent}}$, $\Hat{\mathbf{P}}_{\SI{50}{\percent}}$, and $\Hat{\mathbf{P}}_{\SI{25}{\percent}}$, respectively, computed from the devised network~\eqref{eq6:mlp_structure}.
Notice that, %for the sake of visualization, 
the 3D sound field is depicted in the $\mathrm{xy}$-plane at different $\mathrm{z}$ elevations and time snapshots.
Moreover, the pressure fields of each row are normalized in the range $[-1,1]$.

Inspecting the results, we observe accurate reconstructions of the sound fields for all the different microphone setups considered, as confirmed in Figure~\ref{fig6:pressure_reconstruction_examples}.
As expected, $\Hat{\mathbf{P}}_{\SI{75}{\percent}}$ achieves the best performance with an average $\nmse = \SI{-20.71}{\decibel}$, while the average $\nmse$~\eqref{eq6:nmse_tot} decreases to $\SI{-18.81}{\decibel}$ and $\SI{-13.88}{\decibel}$ when considering $\tilde{\mathcal{M}}_{\SI{50}{\percent}}$ and $\tilde{\mathcal{M}}_{\SI{20}{\percent}}$ of microphones, respectively.
As a matter of fact, the different numbers of observations that sample the acoustic field highly impact the overall reconstructions in the desired 3D region.
However, even starting from $16$ microphones, the reconstruction is close to the ground truth, as can be observed in Figure~\ref{fig6:pressure_reconstruction_examples}d.

\vspace*{5mm}
\noindent \textbf{SOTA comparison}\\
In order to validate the devised methodology also with respect to state-of-the-art approaches, we compare the resulting estimations with two model-based methods for sound field reconstruction.
We compute the acoustic reconstructions in the desired 3D region with the Kernel method~\cite{ueno2018kernel} and the TESM~\cite{antonello2017room}.
Notice that, although recent fully data-driven models for sound field reconstruction have been presented in the literature~\cite{pezzoli2022deep, lluis2020sound}, they focus only on the 2D reconstructions of sound field. %, thus making comparison with the proposed method impossible for 3D scenarios.
For this reason, here we provide comparisons only with respect to model-based approaches.

\begin{figure}[!tb]
    \centering
    \includegraphics[width=\columnwidth]{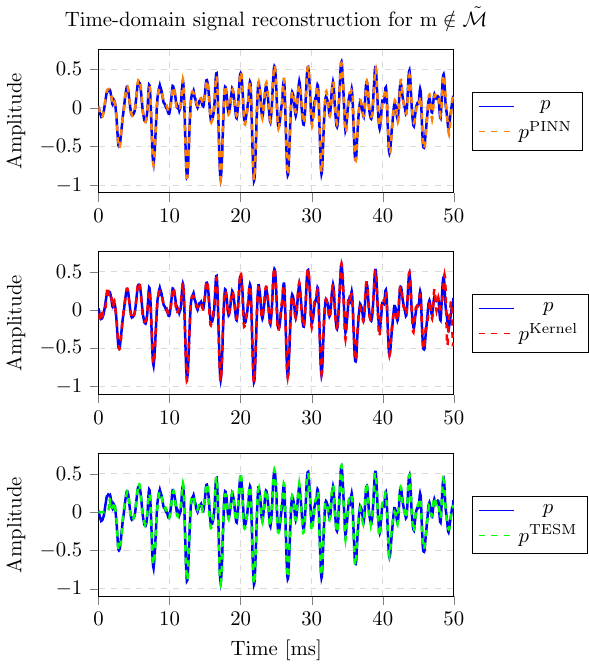}
    \caption{Example of reconstruction comparison of an unknown signal in time-domain. From above to the bottom, the estimate of the proposed method $p^{\mathrm{PINN}}$ and the reconstructions obtained with the Kernel method $p^{\mathrm{Kernel}}$ and TESM $p^{\mathrm{TESM}}$. Each diagram is depicted with the ground truth $p$.}
    \label{fig6:time_reconstruction_comparison}
\end{figure}

The sound field reconstructions in the 3D region have been computed with the Kernel method~\cite{ueno2018kernel} with $\lambda=10^{-3}$ as regularization parameter, sampling rate of $\SI{16000}{\hertz}$, and kernel filter defined in $1025$ points.
On the other hand, differently from the original TESM introduced in~\cite{antonello2017room}, we modified the number and position of the equivalent sources.
Specifically, we arrange $400$ equivalent sources in a sphere with radius $\SI{0.72}{\meter}$. %, thus surrounding the desired acoustic region to model the desired 3D region.
Although defined in frequency-domain and time-domain, respectively, both Kernel interpolation and TESM approaches are constrained to satisfy the physical law of wave propagation~\eqref{eq6:pressure_minimisation_physical}.
In general, they are able to provide good reconstructions of the acoustic fields for each of the considered subsets of observations.

In Figure~\ref{fig6:time_reconstruction_comparison}, we show an example of reconstruction of time-domain signal %for a missing microphone channel 
in a position that does not belong to the available observations, and we compared it with the ground truth pressure $p$.
The estimate of the proposed method $p^{\mathrm{PINN}}$ is depicted above, while the Kernel and TESM reconstruction, denoted as $p^{\mathrm{Kernel}}$ and $p^{\mathrm{TESM}}$, respectively, are below.
In general, the three reconstruction methods fit the ground truth signal.
Notice that, differently from the $p^{\mathrm{PINN}}$, some errors in the initial and final parts of the signal are present for $p^{\mathrm{Kernel}}$ and $p^{\mathrm{TESM}}$.
This is due to the filter in the frequency domain for the kernel method~\cite{ueno2018kernel} and to the causal convolution operator for TESM~\cite{antonello2017room}.

\begin{figure}[!tb]
    \centering
    \includegraphics[width=\columnwidth]{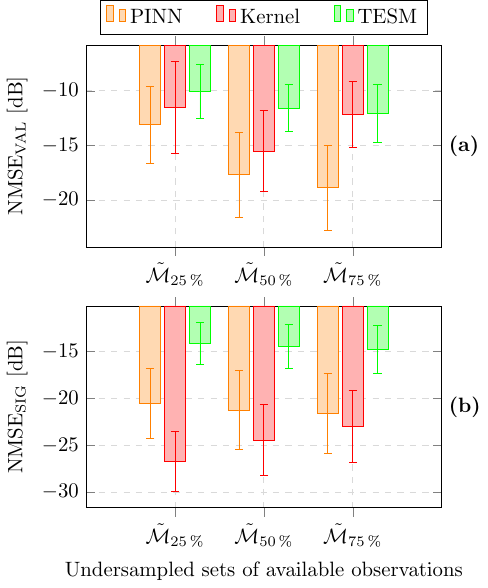}
    \caption{Mean and standard deviation comparison of the $\nmse$ between the proposed PINN approach, the Kernel and TESM methods for different sets of available microphones. (a) the metric considering the unknown pressure signals, (b) the metrics relative to the available observations.}
    \label{fig6:nmse_statistics_comparison}
\end{figure}

To provide a quantitative evaluation of the proposed technique, we compute the reconstruction accuracy for all the considered time windows of the signal.
Figure~\ref{fig6:nmse_statistics_comparison} shows the mean and standard deviation error of the $\nmse$ for the proposed PINN method and the two model-based approaches for the three different observation setups $\tilde{\mathcal{M}}$.
In particular, the $\nmse_{\mathrm{VAL}}$~\eqref{eq6:nmse_val} computed for $\mathrm{m}\notin\tilde{\mathcal{M}}$ and $\nmse_{\mathrm{SIG}}$~\eqref{eq6:nmse_sig} computed for $\mathrm{m}\in\tilde{\mathcal{M}}$ are depicted in the Figure~\ref{fig6:nmse_statistics_comparison}a and Figure~\ref{fig6:nmse_statistics_comparison}b, respectively.

From Figure~\ref{fig6:nmse_statistics_comparison}a, we can notice that the proposed approach and the Kernel method outperform TESM in average for all the configurations. %, both in terms of mean and standard deviation errors.
% [-10.03 -11.6  -12.05] +- [2.46 2.14 2.67]
As a matter of fact, the mean $\nmse_{\mathrm{VAL}}$ for TESM are $\SI{-10.03}{\decibel}$, $\SI{-11.6}{\decibel}$, and $\SI{-12.05}{\decibel}$ for $\tilde{\mathcal{M}}_{\SI{25}{\percent}}$, $\tilde{\mathcal{M}}_{\SI{50}{\percent}}$, and $\tilde{\mathcal{M}}_{\SI{75}{\percent}}$ experiments, respectively.
Although the devised method and the Kernel method achieve similar reconstruction results around $\nmse_{\mathrm{VAL}}=\SI{-12}{\decibel}$ for the case of $\tilde{M} = 16$ microphone observations, the proposed PINN approach reaches better reconstruction performance with a $\nmse_{\mathrm{VAL}}$ difference of $\SI{2.16}{\decibel}$ and $\SI{6.71}{\decibel}$ with respect to $\tilde{\mathcal{M}}_{\SI{50}{\percent}}$ and $\tilde{\mathcal{M}}_{\SI{75}{\percent}}$, respectively.
Moreover, notice that the standard deviation of the proposed PINN and Kernel methods are similar in the three conditions with values around $\SI{3.5}{\decibel}$, $\SI{3.8}{\decibel}$, and $\SI{3.88}{\decibel}$ when moving from $\tilde{\mathcal{M}}_{\SI{75}{\percent}}$ to $\tilde{\mathcal{M}}_{\SI{25}{\percent}}$.

\begin{figure}[!tb]
    \centering
    \includegraphics[width=\columnwidth]{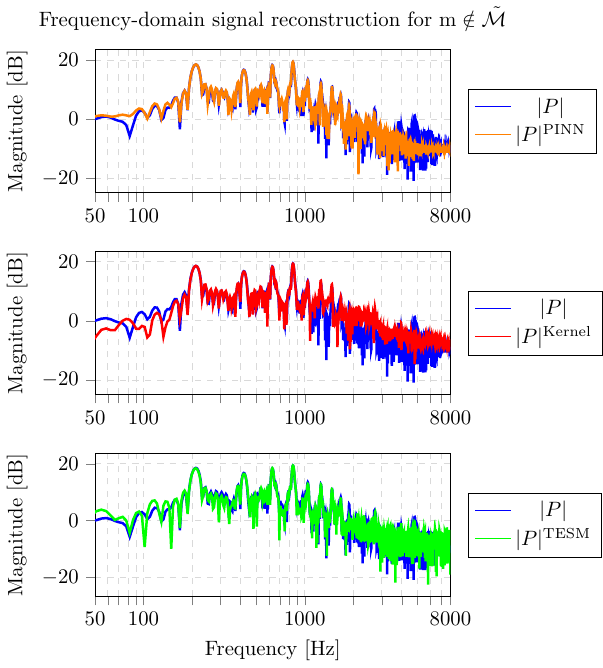}
    \caption{Example of reconstruction comparison of an unknown signal in frequency-domain. From above to the bottom, the magnitude estimate of the proposed method $|P|^{\mathrm{PINN}}$ and the reconstructions obtained with the Kernel method $|P|^{\mathrm{Kernel}}$ and TESM $|P|^{\mathrm{TESM}}$. Each diagram is depicted with the ground truth $|P|$ in $\si{\decibel}$ scale.}
    \label{fig6:frequency_reconstruction_comparison}
\end{figure}

\begin{figure*}[!tb]
    \centering
    \includegraphics[width=\textwidth]{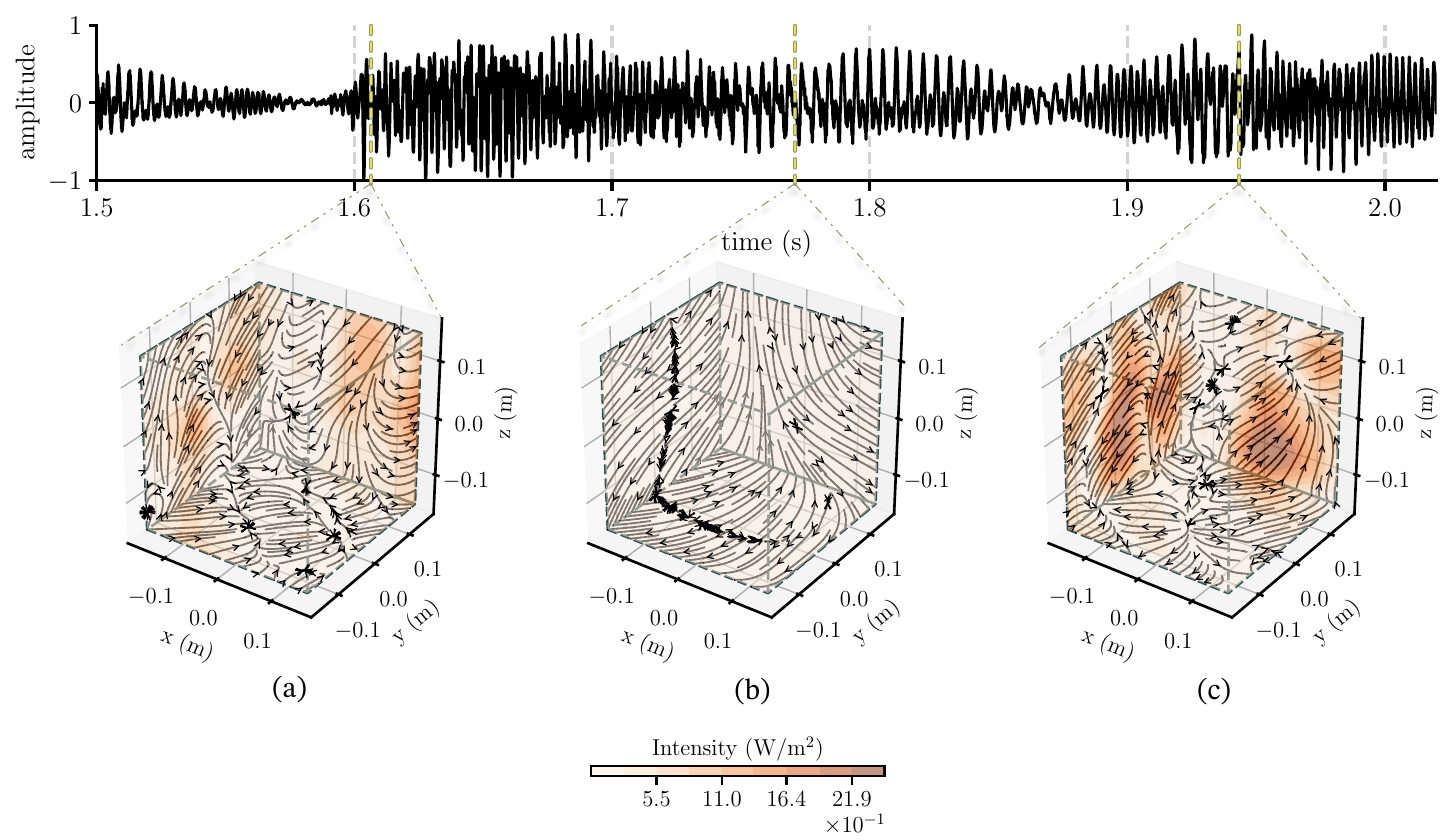}
    \caption{Snapshots of the instantaneous intensity vector at the time (a) $t = 1.612$ s, (b) $t = 1.774$ s and (c) $t = 1.94$ s}
    \label{fig7:intensity_vec}
\end{figure*}

Investigating the results, we believe the similar $\nmse_{\mathrm{VAL}}$ performances for $\tilde{\mathcal{M}}_{\SI{25}{\percent}}$ are due to the relatively small dimension of the 3D region.
With an aperture of $\SI{0.3}{\meter}$ in all three dimensions, the Kernel method is able to provide good interpolations of the available signals~\cite{ueno2018kernel}.
Moreover, inspecting the $\nmse_{\mathrm{SIG}}$ in Figure~\ref{fig6:nmse_statistics_comparison}b, we can notice that the Kernel method retrieves the best fit of the available observations with $\nmse_{\mathrm{SIG}}$ equal to $\SI{-26.72}{\decibel}$, $\SI{-24.46}{\decibel}$, and $\SI{-23}{\decibel}$ for the $1/4$, $1/2$, and $3/4$ of measurements.
% [-26.72 -24.46 -23.  ] +- [3.16 3.8  3.81]
This is due to the design model of the interpolation problem, while the PINN method behaves with a similar trend as the reconstruction for the unknown signals with a $\nmse_{\mathrm{SIG}}$ around $\SI{-21}{\decibel}$.
As a matter of fact, to provide the desired wave equation regularization, the physical term in the loss function of the proposed network~\eqref{eq6:pinn_loss_reconstruction} spreads the error on the whole desired domain without focusing only on the available observations.	

Although the devised PINN method is designed in the time domain, we report in Figure~\ref{fig6:frequency_reconstruction_comparison} a frequency-domain example of the reconstructed signal.
The magnitude of the frequency-domain pressure for the three reconstructions $|P|^{\mathrm{PINN}}$, $|P|^{\mathrm{Kernel}}$, and $|P|^{\mathrm{TESM}}$ are depicted in $\si{\decibel}$ along with the ground truth $|P|$.
We can notice that the proposed method reaches the best accuracy with respect to the two model-based methods.
Although %the peak around 
the trough around $f=\SI{80}{\hertz}$ is not detected, at low frequencies the PINN method is able to match the desired reconstruction somewhat better than the Kernel approach and TESM.
Moreover, the devised solution can capture the frequency contents up to $f=\SI{2500}{\hertz}$, while both Kernel and TESM reconstructions decrease the accuracy after $\SI{1000}{\hertz}$ and $\SI{2000}{\hertz}$, respectively.

\vspace*{5mm}
\noindent \textbf{Intensity field computation}\\
In addition to describing the pressure field, the PINN provides a thorough characterization of the sound field through the reconstruction of intensity flows, governed by Eq. \eqref{eq:insta_intensity}. Illustrated in Fig.~\ref{fig7:intensity_vec}, this visualization incorporates an example of a recorded channel (speech signal) utilized for training, and below is a depiction of these flows. %The trajectories of air particles within the room are delineated by streamlines, with directional arrows indicating their path. Additionally, the contours convey the magnitude of the reconstructed intensity in Watts/m$^2$.

At time instant $t=\SI{1.612}{\second}$ (Fig~\ref{fig7:intensity_vec}a), the snapshot reveals heightened intensity emanating from the speaker's direction $\Vec{r} = (0.15, 0.15, 0.)~\si{\meter}$, along with lateral room reflections (approximately $\Vec{r} = (-0.15, 0.1, 0.08)~\si{\meter}$ and $\Vec{r} = (-0.15, 0, -0.05)~\si{\meter}$). Predominantly, energy is directed from the speaker to the origin and from the rear room corner $\Vec{r} = (-0.15, -0.15, -0.15)~\si{\meter}$.

The subsequent snapshot at time $t=\SI{1.774}{\second}$ (Fig~\ref{fig7:intensity_vec}b) illustrates the phenomenon of destructive interference between oppositely propagating wavefronts. Given that the intensity vector is normal to the propagation direction, energy convergence towards zero along the $\mathrm{y}$ %$=\SI{0}{\meter}$
axis is evident. This observation is further reinforced by the periodic nature of the signal (spoken voice), indicating wavefronts where pressure and velocity are in phase.

Finally, at time $t=\SI{1.94}{\second}$ (Fig~\ref{fig7:intensity_vec}c), substantial lateral energy is observed from all directions, corresponding to reflections propagating in all directions, contributing to a diffuse-like energy distribution. 

%! TEX root=main
\section{Conclusions}\label{sec:conclusions}
In this manuscript, we proposed an approach for the volumetric reconstruction of speech signals with a PINN framework.
We proved the devised architecture to learn an implicit representation of the acoustic field in the target 3D region from a few sparse microphone observations and regularizing the estimates to follow the physical prior knowledge of the acoustic propagation, i.e., the wave equation.

Estimating the time-domain signals of the pressure field, the devised method is able to recover arbitrary acoustic fields, with no constraints on the geometry and acoustic conditions of the environments.
We validated the proposed approach considering a real dataset and considering different subsets of available observations.
The performance has been evaluated in terms of Normalized Mean Square Error between the sound pressure ground truth and the reconstructions.
Moreover, we compared the results with respect to two state-of-the-art methods for the reconstruction of the sound field in the time domain and frequency domain.

We proved %the superiority of 
that our solution outperformed both the baselines preserving the frequency contents of the acoustic field.
Furthermore, to show the potentiality of the proposed PINN method, we presented an example of an intensity field retrieved from the network estimates, by exploiting the automatic differentiation principle.
%Notice that, differently from the other model-based approaches, this can be easily computed exploiting the automatic differentiation principle.

The results prove how the combination of a data-driven approach with the regularization provided by the physical propagation prior of acoustic fields can increase the accuracy of state-of-the-art models, both in the time domain and in the frequency domain.
Nevertheless, we believe the potentiality of such PINN approach in the context of sound field reconstruction can be further exploited for the processing of large 3D regions, such as in rooms or concert halls.
%Therefore, from this preliminary study, we plan to extend the methodology in the near future for applications in the context of acoustic navigation for virtual reality scenarios.

\backmatter

%\bmhead{Supplementary information}
%
%If your article has accompanying supplementary file/s please state so here. 

\section*{Declarations}
\noindent
\textbf{Acknowledgments}\\
%\bmhead{Acknowledgments}
This work was made possible through the support of various entities. Firstly, the European Union contributed under the Italian National Recovery and Resilience Plan (NRRP) of NextGenerationEU partnership regarding ``Telecommunications of the Future" (PE00000001 - program "RESTART"). Additionally, the ``REPERTORIUM project" provided support with grant agreement number 101095065 under Horizon Europe. Cluster II. Culture, Creativity and Inclusive Society, call HORIZON-CL2-2022-HERITAGE-01-02. Lastly, the VILLUM Foundation supported this work under grant number 19179 for the project titled ``Large-scale acoustic holography."\\
%This work was supported by the European Union under the Italian National Recovery and Resilience Plan (NRRP) of NextGenerationEU, partnership on ``Telecommunications of the Future'' (PE00000001 - program ``RESTART''). Additionally, the work was supported by ``REPERTORIUM project. Grant agreement number 101095065. Horizon Europe. Cluster II. Culture, Creativity and Inclusive Society. Call HORIZON-CL2-2022-HERITAGE-01-02''. Finally, this work was supported by the VILLUM Foundation, under grant number 19179, ``Large-scale acoustic holography.''

\noindent
\textbf{Availability of data and materials}\\
This manuscript has no associated data.

\bibliography{bibliography}% common bib file

\end{document}